\documentclass{article}

\usepackage[authoryear,round]{natbib}
\setcitestyle{semicolon,aysep={,}}
\usepackage[hidelinks]{hyperref}
\usepackage{microtype}
\usepackage{graphicx}
\usepackage{subfigure}
\usepackage{booktabs} % for professional tables
\usepackage{amsmath}
\usepackage{array}
\usepackage{hyperref}
\usepackage{tikz}
\usetikzlibrary{arrows.meta,shapes.geometric,positioning}

\usepackage[accepted]{mlsys2025}

\mlsystitlerunning{Guard: Scalable Straggler Detection and Node Health Management for Large-Scale Training}

\begin{document}

\twocolumn[
\mlsystitle{Guard: Scalable Straggler Detection and Node Health Management for Large-Scale Training}

% unless you've provided the [accepted] option to the mlsys2025



% email: 
\mlsyssetsymbol{equal}{*}

\begin{mlsysauthorlist}
\mlsysauthor{Guanliang Liu}{amz}
\mlsysauthor{Abhinandan Patni}{amz}
\mlsysauthor{Congzhu Lin}{amz}
\mlsysauthor{Zoe Zeng}{amz}
\mlsysauthor{Jack Wittmayer}{amz}
\mlsysauthor{Josh Wu}{amz}
\mlsysauthor{Ashvin Nihalani}{amz}
\mlsysauthor{Binxuan Huang}{amz}
\mlsysauthor{Yinghong Liu}{amz}
\mlsysauthor{Rory Na}{amz}
\mlsysauthor{Anthony Ko}{amz}
\mlsysauthor{Alexander Zhipa}{amz}
\mlsysauthor{Cong Cheng}{amz}
\mlsysauthor{Mi Sun}{amz}
\mlsysauthor{Vijay Rajakumar}{amz}
\mlsysauthor{Rejith George Joseph}{amz}
\mlsysauthor{Parthasarathy Govindarajen}{amz}
\end{mlsysauthorlist}

\mlsysaffiliation{amz}{Store Foundational AI, Amazon, Seattle, WA, USA}
\mlsyscorrespondingauthor{Guanliang Liu}{guanlian@Umich.edu}
% \mlsysaffiliation{amz}{Department of Computation, University of Torontoland, Torontoland, Canada}

\mlsyskeywords{Machine Learning, MLSys, Amazon}

\vskip 0.3in

\begin{abstract}
Training frontier-scale foundation models involves coordinating tens of thousands of GPUs over multi-month runs, where even minor performance degradations can accumulate into substantial efficiency losses. Existing health-check mechanisms, such as NCCL tests or GPU burn-in, primarily focus on functional correctness and often fail to detect fail-slow behaviors that silently degrade system performance. In this paper, we present \textbf{Guard}, a scalable system for detecting stragglers and ensuring node health in large-scale training clusters. Guard combines lightweight online performance monitoring during training with an offline node-sweep mechanism that systematically evaluates and qualifies nodes before they participate in production workloads. This design enables Guard to detect both acute failures and long-running fail-slow behaviors that traditional diagnostics cannot capture. Deployed on large-scale foundation model pretraining workloads, Guard improves mean FLOPs utilization by up to \textbf{1.7$\times$}, reduces run-to-run training step variance from \textbf{20\%} to \textbf{1\%}, increases mean time to failure (MTTF), and significantly reduces operational and debugging overhead. These results demonstrate that proactive straggler detection and systematic node qualification are critical for maintaining stable and efficient large-scale training.

\end{abstract}

]

\printAffiliationsAndNotice{}


\section{Introduction}

As large language model (LLM) training has grown in scale from hundreds to thousands to tens of thousands of GPUs, maintaining node health has become a key bottleneck to sustaining high throughput over an extended period of time. \citep{gpipe,pipedream,zero,zerooffload,megatron,megascale,llama3,deepseekv3,gemini15,deepspeed,palm,tensorparallel}. Foundation model training runs now employ hybrid parallelism, combining data, tensor, pipeline, and expert parallelism, across thousands of accelerators that are interconnected by hierarchical high-speed network. \citep{pipedream,zerooffload,megatron,flux,centauri,lowgpuutil,switchtransformer,nvlink}. In such a workload, even a single underperforming node can slow down global training progress as hybrid parallelism significantly increases synchronous NCCL operations such as all-reduce and all-to-all that are bounded by the slowest participant \citep{megascale,flux,centauri,lowgpuutil,nvlink}. While fail-stop faults such as GPU crashes or hardware errors are relatively straightforward to detect and recover from, a more subtle and persistent class of failures known as “grey nodes” have emerged as a major obstacle to stable large-scale training \citep{grayfailure,failslow,dlfailures,superbench,systemimbalance,systemreliability}.

Grey nodes refer to machines that pass standard health checks designed to detect hard failures but exhibit degraded performance during real workloads \citep{grayfailure,failslow,dlfailures,superbench,llmcharacterization,moe,systemimbalance,mlsys2023monitor,thermal}. Unlike fail-stop faults, grey nodes do not immediately cause errors in training jobs; rather, they silently reduce the overall throughput of the job, with their impact accumulating and worsening over time \citep{megascale,llmcharacterization,moe,llama3,tensorparallel,mlsys2023monitor}. In large-scale training runs, a single grey node's impact is magnified as the workload requires thousands of GPUs to synchronize frequently during each training step, causing the slowest node to gate progress\citep{megascale,flux,centauri,lowgpuutil,stabilityai,nvlink}. Existing validation mechanisms such as NCCL tests, GPU burn-in procedures, or short stress benchmarks focus primarily on hardware functionality and communication correctness \citep{dcgm,eud,xid,superbench,stabilityai,nvlink,profiling}. Even with improved on-chip diagnostics reducing silent data corruption and other hardware-level fault rates, these error detection mechanisms only ensure basic operability; performance-impacting faults that grey nodes exhibit  under long-duration, mixed compute–communication workloads persist \citep{dlfailures,flux,centauri,lowgpuutil,llmcharacterization}. Consequently, large production clusters often silently experience prolonged efficiency loss. \citep{grayfailure,failslow,dlfailures,llmcharacterization,moe,systemimbalance,stragglerstudy}.

Our production observations reveal that grey nodes can have numerous root-causes that manifest across multiple system layers. At the hardware level, thermal throttling, degraded high-bandwidth network adapters, and unstable NVLink interconnects can lead to reduced GPU frequency or increased communication latency \citep{megascale,dcgm,eud,xid,stabilityai,profiling}. At the system level, CPU or PCIe bandwidth limitations can further exacerbate these slowdowns 

\citep{flux,centauri,lowgpuutil,nvlink}. Despite these issues being minor in isolation, they are workload- and synchronization-dependent making them difficult to proactively diagnose. When they appear in multi-week training runs, their cumulative effect results in significant inefficiencies. \citep{dlfailures,superbench,llmcharacterization,moe,systemimbalance}. This observation highlights a critical gap in current large-scale training infrastructure: the lack of continuous, non-intrusive, and performance-sensitive node-level health monitoring mechanisms that leverage workload-specific performance metrics. \citep{dcgm,superbench,moe,stabilityai,systemimbalance,hpcc}.

To address these gaps, we present Guard, a system for straggler detection and node health management designed for large-scale hybrid-parallel trainin, as illustrated in Figure~1. Guard integrates online performance monitoring with an offline diagnostic workflow to systematically detect and localize both transient and persistent gray-node behaviors, using training-step time as the primary end-to-end performance signal.

At its core, Guard consists of two tightly coupled components. The online component continuously tracks key hardware and communication metrics—including GPU frequency, temperature, power consumption, network retransmission events, and effective bandwidth—enabling low-overhead detection of anomalous behavior without disrupting ongoing training jobs. These signals provide real-time visibility into performance degradations that are otherwise difficult to observe through traditional health checks.

Complementing the online monitoring layer, Guard includes an offline node-sweep mechanism that executes lightweight diagnostic workloads to reproduce realistic communication and synchronization patterns on a subset of nodes. This design allows Guard to isolate underperforming components, such as degraded interconnects or faulty network interfaces, while minimizing interference with production workloads. Together, the online and offline components form a unified system that enables accurate diagnosis of both transient and persistent performance issues.

\begin{figure*}[t]
\centering
\begin{tikzpicture}[
  font=\small,
  node distance=10mm and 12mm,
  >=Stealth,
  block/.style={rectangle, rounded corners, draw, align=center, minimum height=8mm, minimum width=28mm},
  decision/.style={diamond, draw, aspect=2, align=center, inner sep=1pt},
  line/.style={->, thick}
]

% Nodes
\node[block] (start) {Job starts \\ training};
\node[block, right=of start] (monitor) {Monitor system \\ (real-time node metrics)};
\node[decision, below=of monitor] (bad) {Bad node\\ detected?};
\node[block, below=of bad] (remove) {Remove node from \\ Good Node Pool};
\node[block, below=of remove] (repair) {Repair / Mitigation};
\node[block, below=of repair] (sweep) {Offline Node Sweep};
\node[decision, below=of sweep] (pass) {Sweep passed?};
\node[block, left=15mm of pass] (quarantine) {Keep node \\ quarantined};
\node[block, right=15mm of pass] (return) {Return node to \\ Good Node Pool};

% Edges
\draw[line] (start) -- (monitor);
\draw[line] (monitor) -- (bad);
% \draw[line] (bad) -- node[above,sloped]{No} ++(20mm,0) |- (monitor);
\draw[line] (bad) -- node[right]{Yes} (remove);
\draw[line] (remove) -- (repair);
\draw[line] (repair) -- (sweep);
\draw[line] (sweep) -- (pass);
\draw[line] (pass) -- node[above]{No} (quarantine);
\draw[line] (pass) -- node[above]{Yes} (return);
\draw[line] (return) |- node[pos=0.25, right]{Back to service} (monitor);
\draw[line] (quarantine.west) -| ++(-10mm,25mm) |- node[pos=0.25, left]{Retry later} (repair.west);

\end{tikzpicture}
\caption{Automated node health management workflow. }
\label{fig:node_health_flow}
\end{figure*}
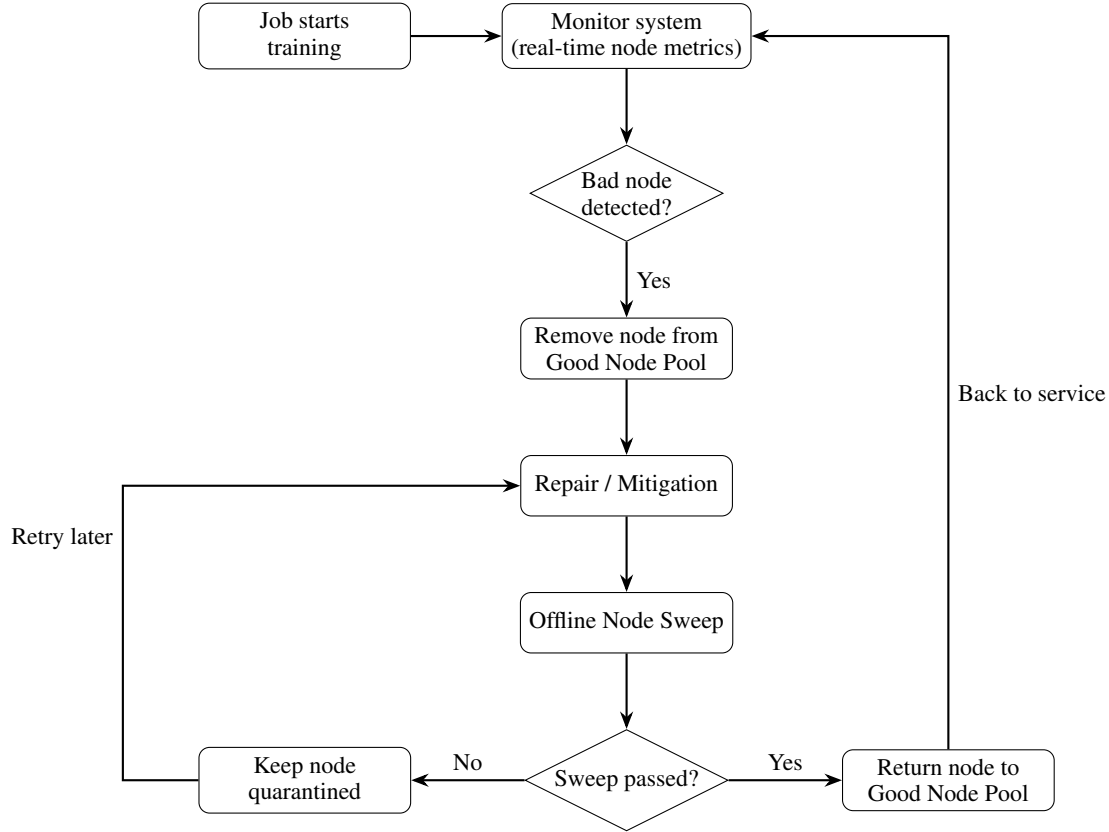

Finally, we present a sample triage workflow that demonstrates how Guard integrates detection, diagnosis, and mitigation into a closed-loop operational pipeline as shown in  Figure~\ref{fig:node_health_flow}. Through this system-level design, Guard provides a practical, scalable, and effective solution for maintaining node health and mitigating stragglers in large-scale distributed training. We have open-sourced a subset of monitoring and detection tools in Amazon’s Foundational Software Kit for AI Training (fkat)\footnote{\url{https://github.com/amzn/fkat}}
.

The remainder of this paper is organized as follows. Section~2 describes our production cluster environment and the motivation for our approach. Section~3 analyzes real-world performance degradation patterns observed in industrial-scale LLM training. Section~4 introduces the online monitoring subsystem, while Section~5 presents the offline node-sweep methodology. Section~6 details the end-to-end triage workflow, and Section~7 evaluates the system through extensive experiments and discusses the broader implications of our findings.

\label{submission}

\section{Background and Motivation}

Detecting and mitigating performance-degraded nodes has become fundamentally more challenging
in modern large-scale training systems due to the combined effects of architectural complexity,
dynamic workloads, and extreme scale.

Modern large-scale language model training relies on increasingly complex hybrid parallelism
strategies—including data, tensor, pipeline, and expert parallelism—that require sustained and
high-volume communication across thousands of GPUs
\citep{gpipe,pipedream,zero,zerooffload,megatron,megascale,flux,centauri,lowgpuutil,switchtransformer,nvlink}.
In such environments, overall training performance is determined not only by the raw computational
capacity of individual accelerators, but by the efficiency of global synchronization across the
entire system.
As a result, a single underperforming node can slow down global progress, since all participants
must wait at synchronization barriers during collective operations making system-wide performance sensitive to even small per-node slowdowns.
\citep{megascale,flux,centauri,lowgpuutil,tensorparallel}.

The architectural complexity of modern foundation models further exacerbates this issue.
In Mixture-of-Experts (MoE) models, expert parallelism distributes subsets of experts to different GPUs, leading to
inherently imbalanced and dynamically changing workloads
\citep{moe,llama3,deepseekv3,gemini15,bloom,tensorparallel}.
Because routing decisions and token distributions vary over time, a degraded node may remain
undetected during initial validation phases, only manifesting as a straggler under specific
communication or workload conditions later in training where its impact compounds across repeated synchronization points
\citep{megascale,llmcharacterization,moe}.

In addition to model-level complexity, large production clusters experience continuous background
fluctuations at both the hardware and system levels.
Thermal variations, power-management adjustments, and transient congestion in inter-node communication
paths can degrade performance without triggering explicit hardware alarms
\citep{dlfailures,dcgm,eud,xid,superbench,stabilityai,profiling}.
These soft degradations may persist for extended periods, silently eroding training throughput in specific workloads and thereby
remaining invisible to conventional monitoring mechanisms
\citep{superbench,llmcharacterization,systemimbalance}.

Conventional node validation techniques—such as short-duration communication benchmarks or GPU
burn-in diagnostics—are primarily designed to verify functional correctness under idealized
conditions
\citep{dcgm,eud,xid,stabilityai,nvlink}.
However, these tests fail to capture performance degradations that are workload-dependent,
intermittent, or only observable under long-running, mixed compute–communication workloads
\citep{flux,centauri,lowgpuutil,llmcharacterization,hpcc}.
As a result, nodes that pass standard validation may later exhibit degraded behavior when exposed
to realistic training traffic.

As training scales further, the operational impact of such undetected inefficiencies becomes
substantial.
Even modest per-node performance deviations on the order of a few percent can accumulate across
thousands of GPUs and millions of training steps, translating into significant increases in
time-to-solution and infrastructure cost
\citep{megascale,llmcharacterization,moe,llama3}.
Maintaining a consistently healthy cluster therefore requires mechanisms that go beyond detecting
fail-stop failures, and instead continuously identify and mitigate silent or gradual performance
degradation at scale
\citep{grayfailure,failslow,dlfailures,systemimbalance,hpcc,systemreliability}.

\section{Job Slowdown Analysis}
\label{sec:slowdown_analysis}

Large-scale distributed training frequently suffers from performance degradations that do not
cause job failures but significantly reduce throughput.
In synchronization-heavy workloads, overall progress is bounded by the slowest participants,
causing even small node-level slowdowns to be amplified at the job level.
Unlike fail-stop faults, these degradations often evade traditional health checks and persist
silently for long periods, making them particularly costly in multi-week training runs.

Across production workloads, we observe two recurring slowdown behaviors.
First, a subset of nodes consistently exhibits slower performance than the cluster baseline,
leading to a persistent inflation of training step time.
Second, transient step-time spikes appear intermittently, often correlated with communication
instability or resource contention.
While these spikes may seem minor in isolation, they compound significantly at scale and can
trigger cascading slowdowns or timeouts in large hybrid-parallel jobs.

\begin{figure}[t]
  \centering
  \includegraphics[width=\linewidth]{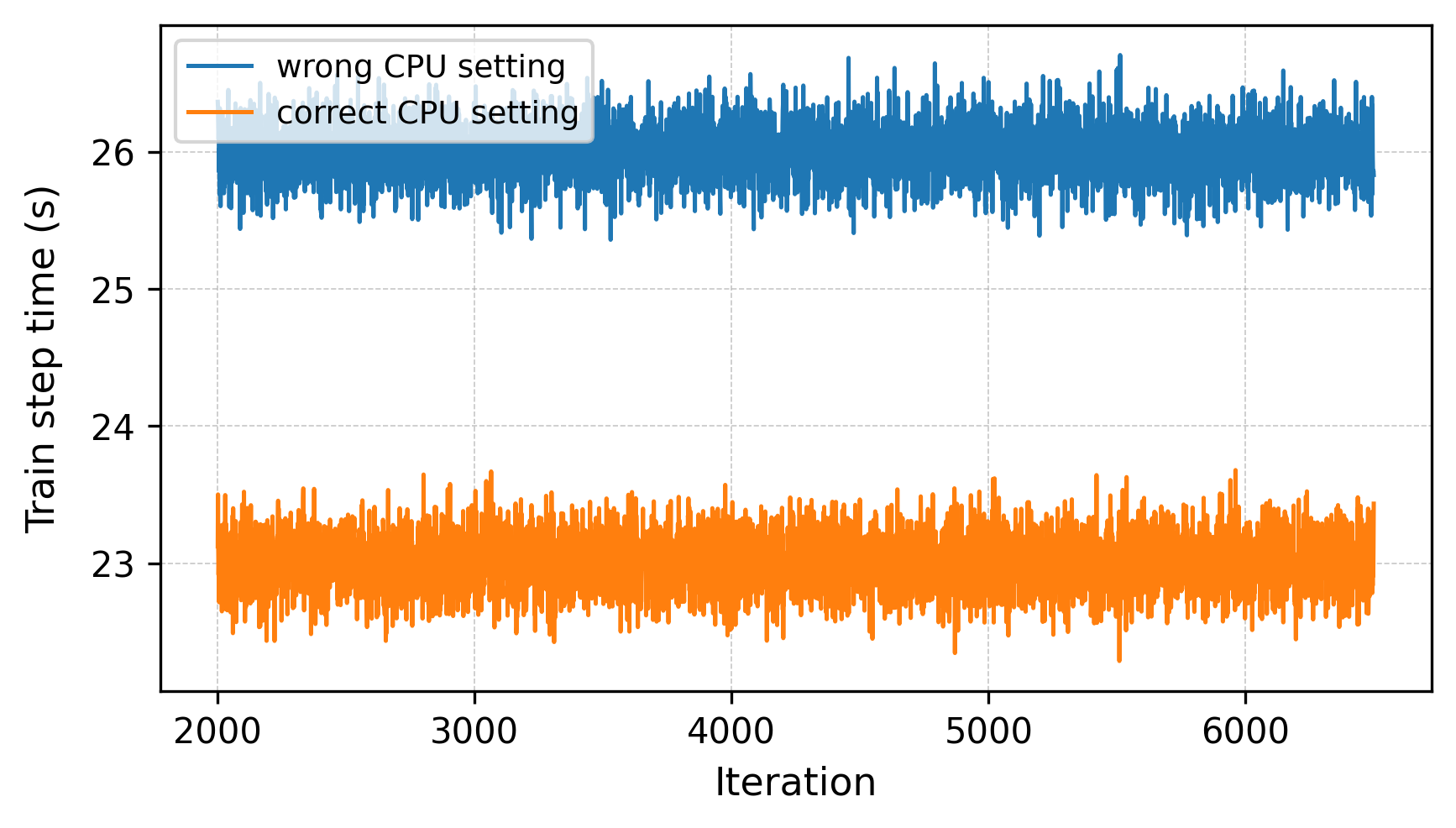}
  \caption{Performance differences under different CPU settings}
  \label{fig:cpu_time_diff}
\end{figure}

Our investigation reveals that these slowdowns arise from multiple system layers.
On the CPU side, insufficient allocation or bandwidth limitations delay data movement and
communication coordination.
Figure~\ref{fig:cpu_time_diff} shows that the correct CPU setting can increase training speed up to 15\% even when GPU hardware and utilization remain unchanged.
This highlights that GPU utilization alone is insufficient for assessing node-level performance.

% \begin{figure}[t]
%   \centering
%   \includegraphics[width=\linewidth]{train_step_time_diff.png}
%   \caption{different CPU settings' training step time}
%   \label{fig:cpu_time_diff}
% \end{figure}

Communication-related issues further exacerbate slowdown behavior.
When a high-bandwidth interconnect adapter degrades or becomes unavailable, modern communication
libraries such as NCCL may transparently reroute traffic through alternative paths to preserve
functional correctness.
Although this avoids immediate job failure, it reduces effective bandwidth and introduces routing
asymmetry.
As shown in Figure~\ref{fig:train_step_time}, resolving such degraded communication paths reduces
training step time from 8.7\,s to 8.4\,s.
In large-scale or MoE workloads with repeated synchronization phases, these sub-second differences
accumulate into substantial efficiency loss.

% \begin{figure}[t]
%   \centering
%   \includegraphics[width=0.8\linewidth]{8.7s_8.4s.png}
%   \caption{Training step time reduction after removing a degraded communication path.
%   Small per-step differences compound significantly at scale.}
%   \label{fig:train_step_time}
% \end{figure}

GPU-side performance degradation represents another major contributor.
Thermal throttling, power delivery instability, and aging hardware can reduce sustained clock
frequency or effective throughput without triggering explicit errors.
A single underperforming GPU can slow down collective operations for its entire parallel group,
propagating localized degradation into cluster-wide straggler behavior.

These issues are rarely detected by traditional node validation methods such as GPU burn-in tests
or short NCCL communication checks.
Burn-in tests emphasize functional correctness over sustained throughput, while NCCL tests prioritize
connectivity and tolerate performance loss through transparent rerouting.
As a result, nodes that pass standard validation may still behave as grey nodes in production,
consistently lagging behind peers and degrading overall training efficiency.

Together, these observations motivate the need for continuous, performance-aware monitoring during
training and independent offline verification before nodes are returned to service.
The following sections introduce the online node health monitoring system that detects performance anomalies in real time. Section~\ref{sec:offline_sweep} introduces an offline node sweep mechanism that validates
node health under realistic workloads 

\subsection{CPU Setting Limitations}

CPU configuration plays a critical role in large-scale training performance and involves two key aspects: CPU allocation per GPU worker and CPU frequency settings. The optimal CPU allocation is highly workload-dependent, varying across training paradigms such as dense models and Mixture-of-Experts (MoE), while overly aggressive frequency tuning can negatively impact overall system stability.

In large-scale foundation model training, CPUs are responsible for several performance-critical tasks, including data loading, checkpointing, and coordinating intra-node GPU communication. As training scales, CPU bandwidth increasingly becomes a limiting factor. Once CPU resources approach their bandwidth limits, these essential operations can become bottlenecked, leading to cascading performance degradation and reduced overall training throughput.

Specifically, CPU bandwidth limitations can affect:
\begin{itemize}
    \item \textbf{Data Loading}: Slower preprocessing and delivery of training samples to GPUs;
    \item \textbf{Checkpoint Operations}: Increased latency in saving and restoring model states;
    \item \textbf{Inter-GPU Communication}: Reduced efficiency in coordinating and transferring data among GPUs.
\end{itemize}

In our production experiments, we identify two key CPU configuration parameters that significantly impact training performance. The first is CPU frequency. By default, most servers enable dynamic frequency scaling; however, we observe that disabling dynamic scaling and using a fixed CPU frequency can lead to higher and more stable training throughput. The second parameter is the number of CPU cores allocated to each training pod. This requirement is highly model-dependent: models with heavier communication patterns, such as Mixture-of-Experts (MoE), require substantially more CPU resources than dense models to achieve optimal performance.

When the combined workload exceeds CPU bandwidth capacity, these operations experience increased latency, directly impacting the overall training throughput. This bottleneck becomes particularly pronounced in distributed training scenarios where efficient communication and data movement are crucial for maintaining training efficiency. The detailed results are shown in Figure~\ref{fig:cpu_time_diff}. Without proper CPU settings, training throughput can decrease by up to 15\%.

\subsection{Slow communication}
Communication inefficiencies represent another critical challenge in production training environments. While the current NCCL routing architecture assigns dedicated network adapters to each GPU, the system's fallback mechanism can mask underlying hardware issues. When a GPU's network adapter fails, NCCL automatically reroutes communication through another GPU's adapter (typically GPU 0) without triggering hardware failure alerts. For instance, if GPU 5's adapter malfunctions, its communication will be redirected through GPU 0's network connection.

This fallback mechanism, while preventing immediate job failures, introduces significant performance degradation across the training cluster. The impact is particularly severe in Mixture of Experts (MoE) model training, where communication patterns are more complex. MoE architectures require two critical synchronization points: token dispatch and result combination, both involving communication across the entire Expert parallelism group. When a node operates sub-optimally, it creates bottlenecks at these synchronization points. 

The performance impact compounds significantly in modern architectures. For example, in a model with 32 MoE layers, these communication inefficiencies accumulate across each layer, resulting in substantial overall training slowdown. This cascading effect demonstrates why detecting and addressing sub-optimal communication patterns is crucial for maintaining training efficiency at scale.
\begin{table}[t]
\centering
\caption{Comparison between normal and abnormal GPU--NIC mappings.}
\label{tab:ib_eth_examples}
\small
\renewcommand{\arraystretch}{1.2}
\begin{tabular}{l p{2.2cm} p{2.2cm}}
\toprule
 & \textbf{GPU0 (Normal)} & \textbf{GPU7 (Abnormal)} \\
\midrule
Expected NIC & 0 & 7 \\
Actual NIC   & 0 & 0 (misrouted) \\
Status       & Normal & Faulty (adapter down) \\
\bottomrule
\end{tabular}
\end{table}

\begin{figure}[t]
    \centering

    \includegraphics[width=0.8\linewidth]{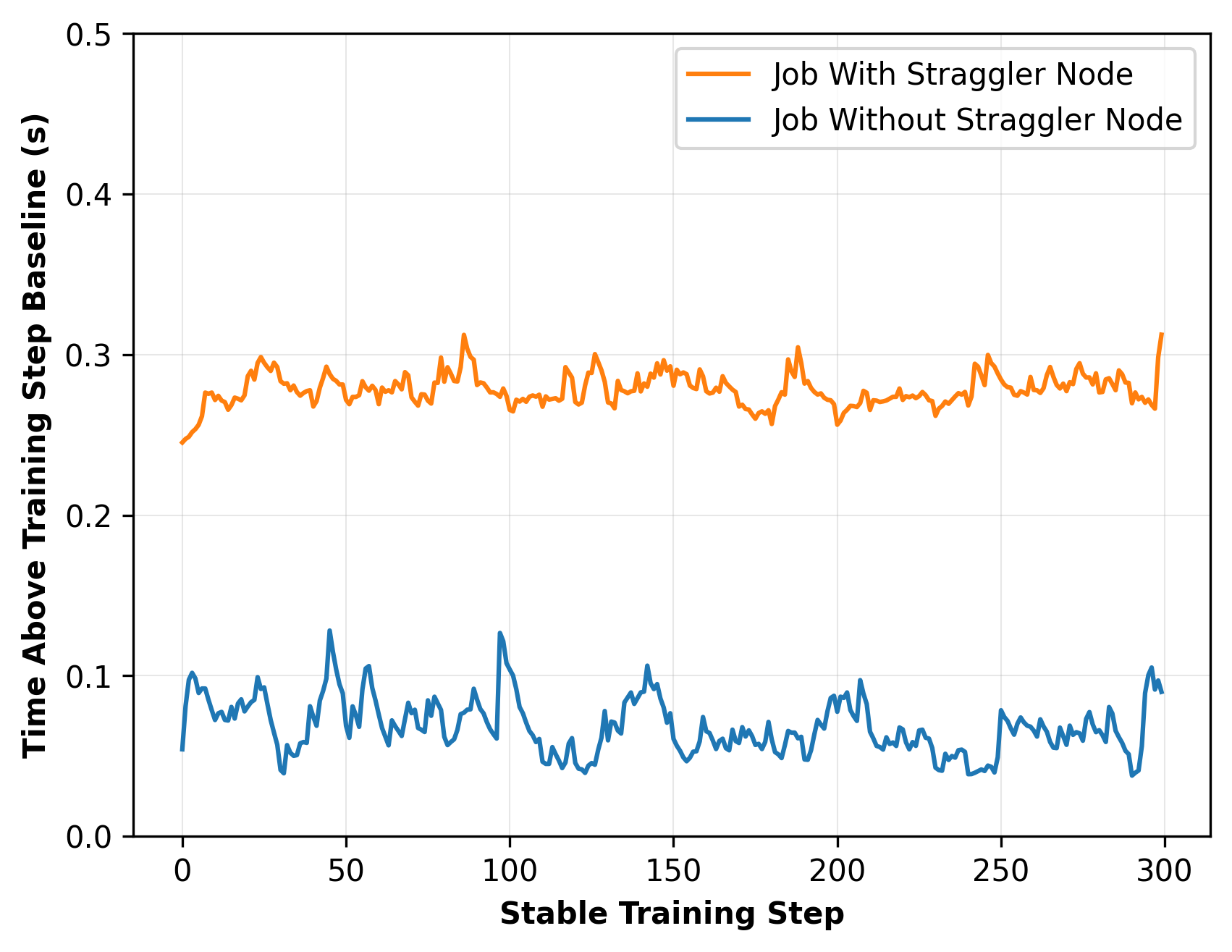}
    \caption{Training step time reduction from 8.7s to 8.4s.}
    \label{fig:train_step_time}

    \vspace{2pt} % 控制上下图之间的间距，数值可微调 -2pt ~ -8pt
    % 第一张图
    \includegraphics[width=0.8\linewidth]{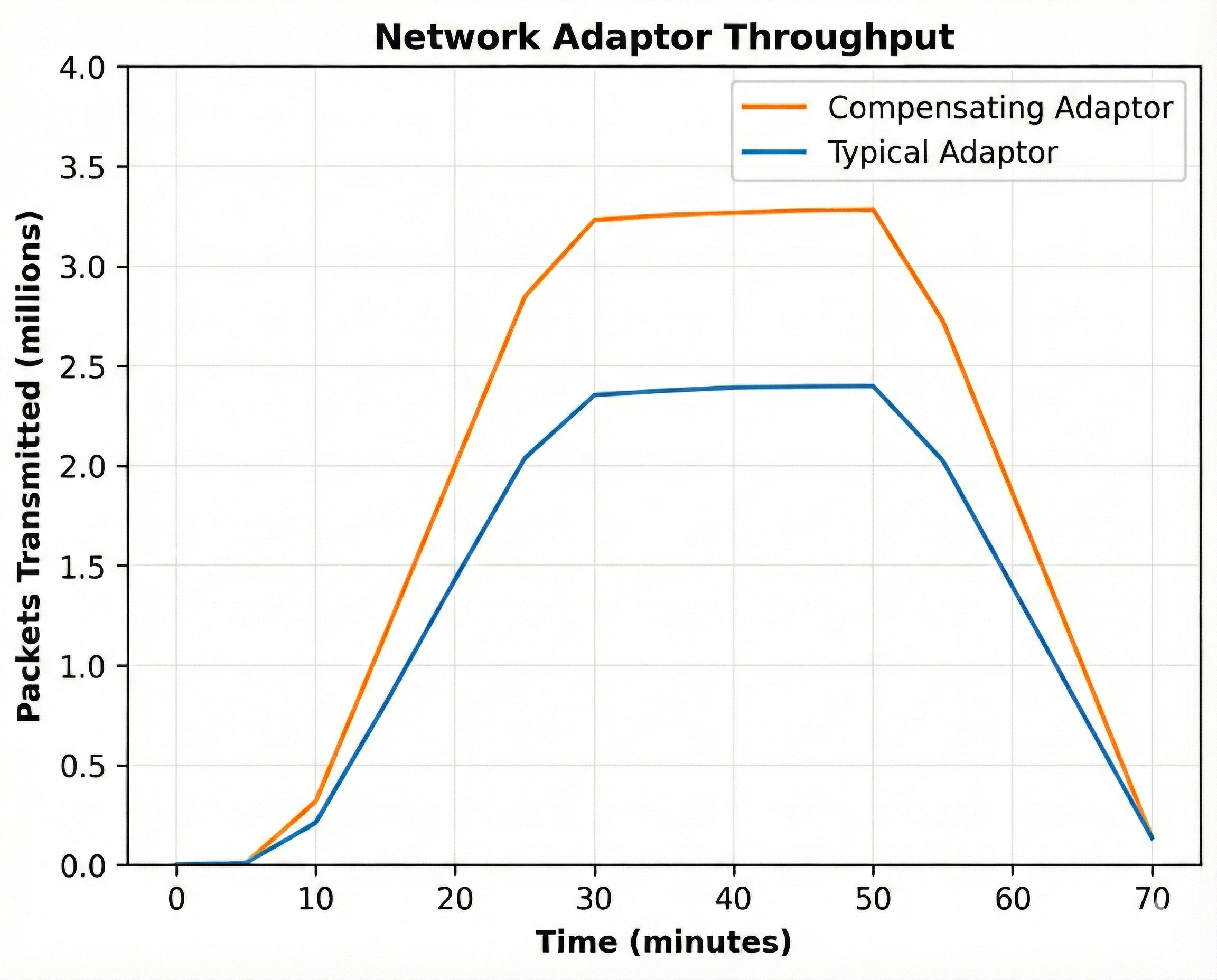}
    \caption{Abnormal network packets transmitted metric.}
    \label{fig:abnormal_ib}

    % 第二张图

\end{figure}

Table 1 shows an abnormal network status. Even when one network adapter is down, the NCCL library automatically reroutes communication through another available adapter to maintain functionality. However, this failover leads to abnormally high workloads on the remaining network adapter, effectively doubling the network traffic as illustrated in Figure ~\ref{fig:abnormal_ib}. Consequently, each training step becomes slower by approximately 0.3 seconds, as shown in Figure~\ref{fig:train_step_time}

\subsection{GPU Calculation Performance Degradation}

GPU performance degradation represents one of the most common yet subtle causes of node-level slowdowns in large-scale distributed training. In production environments, multiple hardware and environmental factors, such as thermal throttling, unstable power delivery, and component aging, can lead to sustained performance loss on individual GPUs even when the node passes standard validation tests.

Thermal throttling is a primary contributor to GPU performance degradation. During intensive and prolonged training workloads, GPUs operate near their thermal design limits. When the temperature exceeds predefined thresholds, the hardware automatically reduces core clock frequency or power draw to prevent overheating. These protective mechanisms, while preserving device safety, directly reduce computational throughput. As shown in Table \ref{sample-table}, an increase in GPU temperature from 50°C to 77°C can result in a core frequency drop from 1.93\,GHz to 1.38\,GHz, leading to significant slowdowns during synchronization-heavy training steps.

In addition to temperature effects, power delivery instability is another critical factor influencing GPU performance. Modern GPUs rely on consistent and high-quality power supply to sustain peak clock frequencies. Voltage fluctuations, current limits imposed by faulty power distribution units (PDUs), or degraded power cables can cause the GPU to enter conservative performance states. In our production observations, nodes with slightly lower power draw (10--15\% below cluster average) exhibited noticeably reduced FLOPS utilization despite showing normal utilization rates and frequencies, suggesting that power constraints can silently limit performance.

Memory subsystem efficiency also plays an important role. Aging hardware or marginal memory modules may experience increased access latency or ECC corrections, reducing effective memory bandwidth and causing stalls in compute pipelines. These issues are rarely detected by burn-in tests, which typically run for short durations and emphasize arithmetic correctness rather than sustained data movement.

Moreover, environmental and mechanical factors—such as dust accumulation, uneven airflow across chassis, or partially degraded cooling fans—can create persistent temperature imbalances between GPUs within the same node. Even a single GPU operating below its nominal frequency can slow down collective operations like AllReduce and AllGather, propagating the slowdown to all participating devices in the cluster.

\begin{table}[t]
  \caption{Relationship between GPU temperature and clock frequency. Higher temperatures trigger protective downclocking, reducing effective throughput.}
  \label{sample-table}
  \vskip 0.15in
  \centering
  \begin{small}
  \begin{sc}
  \begin{tabular}{ccc}
    \toprule
    \textbf{GPU} & \textbf{Temperature (°C)} & \textbf{Core Frequency (GHz)} \\
    \midrule
    GPU0 & 50 & 1.93 \\
    GPU1 & 60 & 1.93 \\
    GPU2 & 69 & 1.78 \\
    GPU3 & 77 & 1.38 \\
    \bottomrule
  \end{tabular}
  \end{sc}
  \end{small}
  \vskip -0.1in
\end{table}

In summary, GPU computation degradation can arise from a combination of thermal, electrical, and memory-level inefficiencies. While each factor may appear minor in isolation, their cumulative effects can cause severe straggler behavior in distributed training. Continuous monitoring of temperature, power draw, frequency stability, and memory error rates is therefore essential for early detection and mitigation of GPU-level degradation before it impacts cluster-wide performance.

\section{Online Node Health Monitoring System}
\label{sec:online_monitor}

Traditional node health checks in large-scale training systems focus primarily
on functional correctness, such as GPU availability or binary link status.
However, our production experience shows that many of the most damaging
performance issues do not cause explicit failures.
Instead, nodes remain operational while silently delivering reduced effective
throughput, becoming stragglers in synchronization-heavy workloads.

To address this gap, we design an online node health monitoring system that
continuously tracks performance-correlated hardware and network signals during
real training runs.
Rather than relying on short validation tests or static thresholds, the system
is grounded in degradation patterns repeatedly observed in production.

\subsection{Proposed Metrics for Continuous Node Health Monitoring}

Based on observations from our production-scale training environment, we identify several hardware and system-level indicators that strongly correlate with node stability and training efficiency. These metrics are continuously collected from all active nodes during training and jointly analyzed to detect anomalies that traditional validation methods fail to capture.

\textbf{GPU Temperature.}
Sustained high GPU temperatures can trigger thermal throttling mechanisms, directly reducing computational performance. While short burn-in tests may not expose this behavior, long-running workloads often reveal persistent thermal imbalances due to cooling inefficiencies. Continuous temperature monitoring allows early detection of thermal anomalies, preventing long-term degradation in sustained workloads.

\textbf{GPU Utilization.}
GPU utilization represents how effectively GPU compute resources are being used. Persistently low utilization can indicate CPU–GPU imbalance, software inefficiency, or internal bottlenecks such as limited memory bandwidth. Monitoring utilization trends helps identify nodes underperforming relative to cluster averages.

\textbf{GPU Clock Frequency.}
Clock frequency serves as a direct indicator of GPU performance state. Frequency drops caused by thermal throttling, power capping, or hardware aging can silently reduce node throughput. Tracking per-GPU frequency stability in real time enables detection of subtle slowdowns that are invisible in coarse-grained cluster metrics.

\textbf{GPU Power Draw.}
Power draw provides an indirect measure of GPU workload intensity. Under normal training loads, stable power consumption is expected. Deviations—especially low power draw despite high utilization—often suggest voltage regulation issues or failing hardware components. In our production observations, nodes showing normal utilization but abnormally low power draw were consistently correlated with reduced FLOPS utilization, validating this as a strong health signal.

\textbf{Network device Error Count.}
Network device error metrics, including packet retransmissions, link retries, and link flaps, reveal low-level network instability. These transient issues rarely appear in short validation tests but can cause synchronization stalls or communication timeouts in long-running distributed training. Monitoring error counts helps identify potential cable degradation or malfunctioning network adapters early.

\textbf{Network device Transmission Rate (Gb/s).}
Transmission rate reflects the effective throughput of network communication. Persistent reductions in transmission rate—without corresponding error spikes—may indicate degraded cables, misconfigured MTU, or thermal throttling of the network interface. This metric provides a holistic view of inter-node communication efficiency.

\textbf{Network device Status.}
Network device may intermittently reset due to hardware faults or driver-level issues. Although NCCL automatically reroutes traffic to maintain connectivity, such fallback paths usually introduce additional latency or bandwidth reduction. Continuous monitoring of link status allows operators to proactively identify and isolate affected nodes before performance degradation propagates.

\subsection{Online Monitoring Strategy}

Motivated by the observations above, our online monitoring system continuously tracks a compact yet informative set of hardware- and network-level metrics that directly reflect effective training performance. These metrics include GPU temperature, clock frequency, power draw, interconnect error counters, and effective communication throughput. Rather than relying on fixed absolute thresholds, all metrics are evaluated relative to peer nodes participating in the same training job, allowing the system to naturally adapt to workload characteristics and hardware heterogeneity.

To ensure robustness, nodes are flagged only when multiple performance indicators exhibit sustained deviation from the peer baseline across consecutive evaluation windows. This multi-signal and temporal filtering strategy effectively suppresses false positives caused by transient fluctuations, while still enabling early detection of persistent performance degradation. Once a node is flagged, it is removed from the healthy node pool and scheduled for offline verification, ensuring that online monitoring remains lightweight and non-intrusive. Final validation is performed through the offline node sweep mechanism described in Section 5.

Importantly, our system does not classify nodes as faulty based solely on hardware error counters. Instead, training step time—representing user-visible performance—is treated as the primary signal, with hardware metrics serving as supporting indicators. Based on the severity of observed performance impact, we apply a tiered response policy that balances mitigation urgency with operational disruption:

\begin{itemize}
    \item \textbf{No observable impact.} If training throughput remains unaffected, the node is marked as \emph{pending verification}. The running job is left unchanged, and the node continues to participate while being monitored more closely.
    \item \textbf{Moderate, sustained slowdown ($\sim$10\%).} When a measurable but tolerable slowdown is detected, the issue is considered actionable but non-urgent. Mitigation is deferred to the next checkpoint to confirm the diagnosis while avoiding unnecessary job interruption.
    \item \textbf{Severe degradation or stalls ($\geq$20\%).} When performance degradation is substantial or progress stalls, the node is deemed harmful to training. The job is immediately restarted with a healthy replacement node, and the affected node is removed from service for remediation.
\end{itemize}

This tiered strategy prevents overreaction to transient anomalies while ensuring timely intervention when performance degradation materially impacts large-scale training efficiency.

% Section~\ref{sec:offline_sweep}.

\section{Offline Node Health Verification Method}
\label{sec:offline_sweep}

\subsection{Why Traditional Node Health Checks Are Insufficient}

Traditional node validation methods such as GPU burn-in tests and NCCL communication tests are
widely used in large-scale distributed training systems to verify basic hardware functionality
before admitting nodes into production clusters.
However, these methods are primarily designed to detect hard failures and short-term instability,
rather than sustained performance degradation under realistic training workloads.

GPU burn-in tests focus on exercising compute units using synthetic arithmetic and memory-intensive
kernels over short durations.
While effective at detecting faults such as overheating or defective cores, they do not sufficiently
stress performance-critical subsystems including memory bandwidth, NVLink interconnects, and PCIe
paths.
As a result, nodes may pass burn-in validation while still exhibiting degraded intra-node
communication or asymmetric GPU performance under sustained load.

Similarly, NCCL tests validate the functional correctness of collective communication primitives
(e.g., AllReduce, Broadcast) but typically run for only a few seconds under idealized
conditions.
Moreover, NCCL’s fault-tolerance mechanisms can transparently reroute traffic around degraded links,
allowing tests to pass while masking reduced bandwidth, routing asymmetry, or network-level
instability.
Such effects often emerge only during prolonged, synchronization-heavy distributed training.

In practice, these limitations lead to the emergence of \emph{grey nodes}: machines that do not fail
outright but consistently lag behind others, inflating step time and reducing overall cluster
throughput.
To address this gap, we adopt a closed-loop node health management framework that combines continuous
online monitoring with an offline verification step.
Online monitoring conservatively detects suspicious behavior during real workloads, after which the
node is removed from the healthy pool.
An offline \emph{node sweep job} is then triggered to perform end-to-end performance validation before
the node is allowed to re-enter production, ensuring both safety and diagnostic accuracy.

\subsection{Single-Node Sweep: Intra-Node Performance Validation}

The single-node sweep targets performance degradations within a single node that frequently evade
traditional validation.
It is designed to expose sustained throughput loss and communication asymmetry while remaining
lightweight enough to run independently of full training jobs.

Specifically, the sweep evaluates:
\begin{itemize}
    \item \textbf{Per-GPU compute throughput}, ensuring consistent sustained FLOPS across all GPUs.
    \item \textbf{Intra-node interconnect bandwidth}, validating NVLink connectivity and symmetry
    through pairwise GPU communication tests.
\end{itemize}

Unlike conventional burn-in tests that emphasize functional correctness, the single-node sweep
explicitly measures sustained throughput and communication efficiency.
This design makes subtle degradations—such as uneven GPU throttling or partially degraded NVLink
paths—directly observable.

\begin{figure}[!t]
    \centering
    \includegraphics[width=0.8\linewidth]{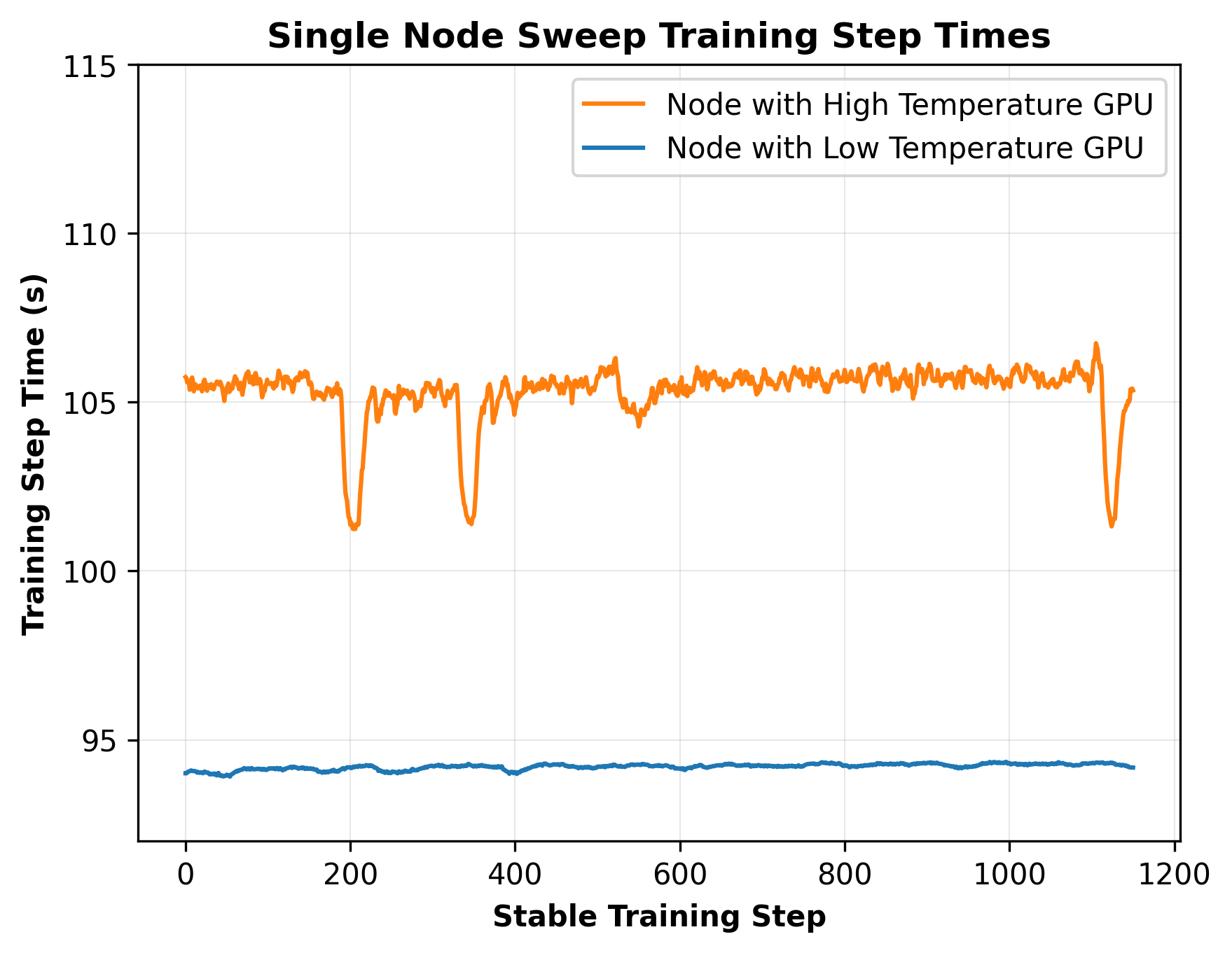}
    \caption{Single-node sweep results showing intra-node performance divergence across GPUs that
    pass traditional validation.}
    \label{fig:single_node_sweep}
\end{figure}

Figure~\ref{fig:single_node_sweep} shows representative results in which GPUs within the same node
exhibit measurable performance divergence despite passing burn-in tests.
Such asymmetries often amplify during collective operations, leading to persistent straggler behavior
in distributed training.

\subsection{Multi-Node Sweep: Inter-Node Communication Validation}

While single-node validation captures local degradation, many stragglers originate from inter-node
communication issues, including degraded high-bandwidth network links or transparent rerouting by the
communication library.
To detect these failures, we perform a multi-node sweep that stresses cross-node collective
communication under controlled conditions.

We evaluate configurations with 2, 4, and 8 nodes and find that most communication-related
degradations are already detectable in the 2-node setting.
Degraded links or misrouted traffic manifest immediately as elevated latency or reduced bandwidth,
making minimal-scale sweeps surprisingly effective.
Although larger configurations increase sensitivity, they offer diminishing returns relative to
their cost.

\begin{figure}[!t]
    \centering
    \includegraphics[width=0.8\linewidth]{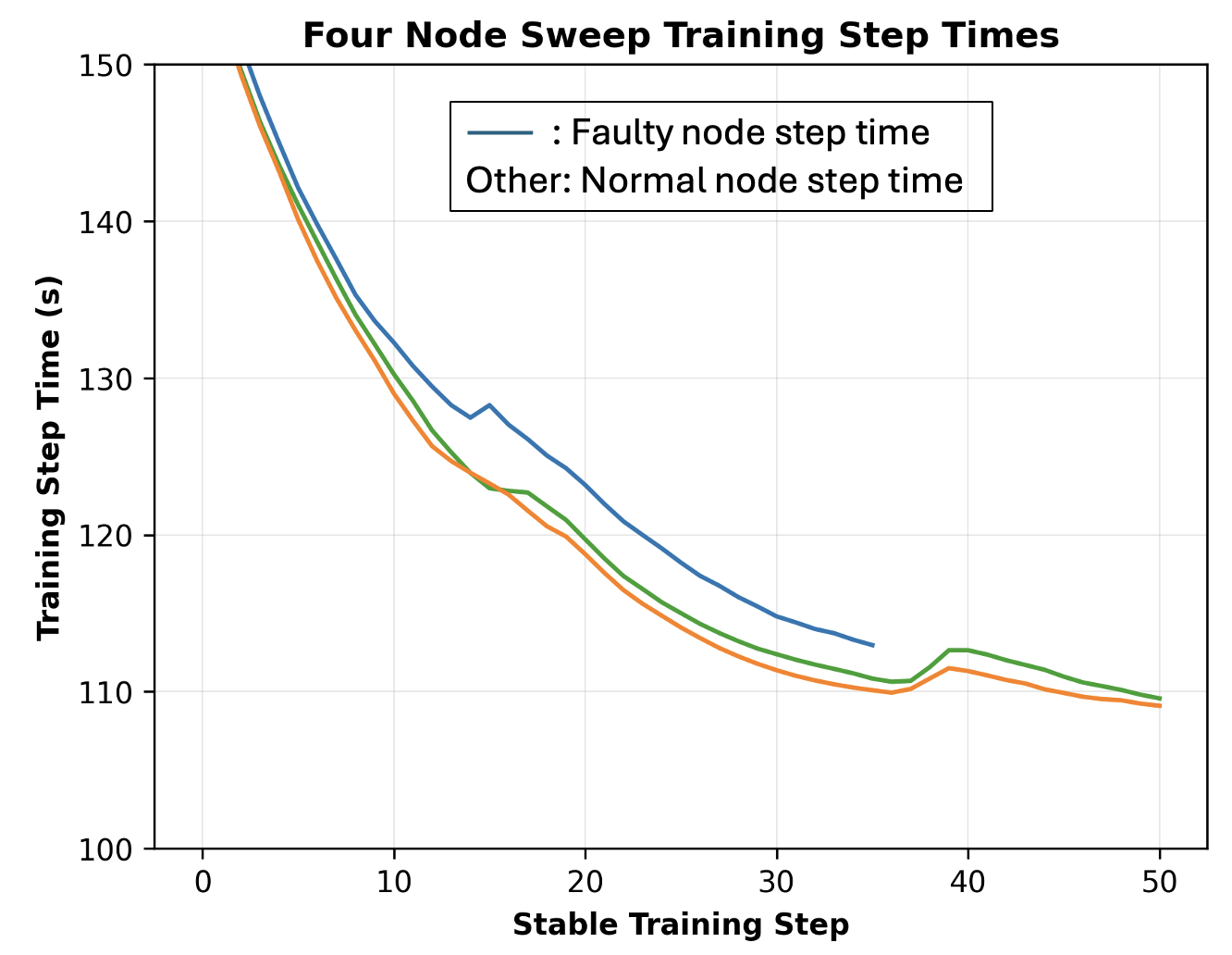}
    \caption{Multi-node (2-node) sweep showing step-time inflation caused by inter-node communication
    degradation.}
    \label{fig:4_node_sweep}
\end{figure}

\begin{figure}[!b]
    \centering
    \includegraphics[width=0.8\linewidth]{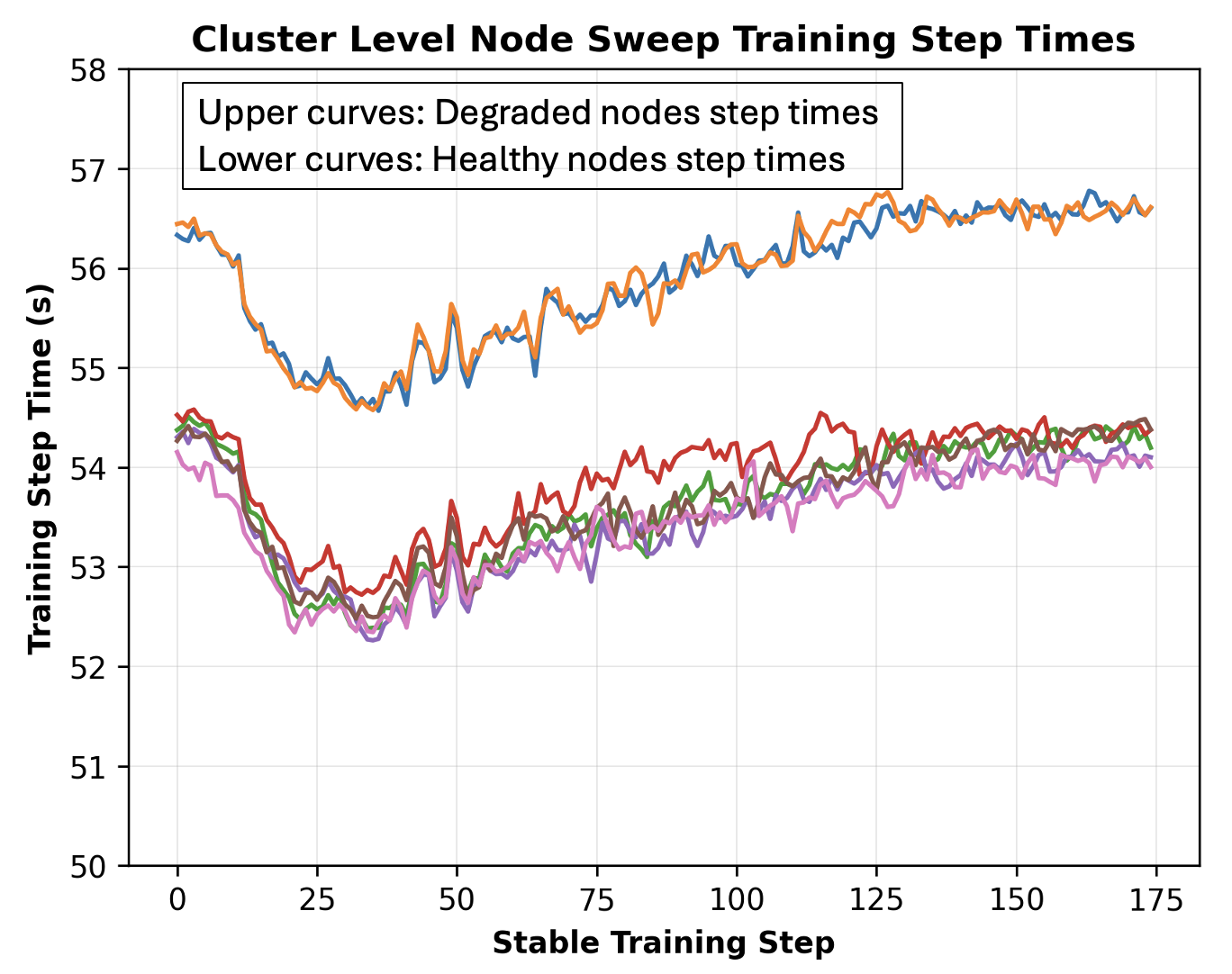}
    \caption{Cluster-level node sweep demonstrating scalability of the offline validation approach
    as faulty nodes are introduced.}
    \label{fig:cluster_node_sweep}
\end{figure}

Figure~\ref{fig:4_node_sweep} illustrates how including a faulty node in a 2-node sweep leads to
consistent step-time inflation.
This behavior scales predictably as more nodes are added, as shown at the cluster level in
Figure~\ref{fig:cluster_node_sweep}, motivating our default use of 2-node sweeps as a cost-effective
verification mechanism.

\subsection{Node Sweep Results and Operational Use}

The node sweep evaluates end-to-end performance under sustained load, succeeding where traditional
NCCL tests fail.
Because it does not rely on short-lived correctness checks, the sweep makes bandwidth loss, routing
asymmetry, thermal throttling, and intermittent network instability directly observable.

Sweep results are interpreted conservatively.
Nodes that pass both single-node and multi-node sweeps are returned to the healthy pool, while nodes
that fail remain quarantined for further repair or re-imaging.
In practice, a sweep duration of 1--2 hours is sufficient to detect persistent compute and
communication degradation.

Rather than periodically sweeping all nodes, we adopt an event-driven strategy in which sweeps are
triggered by anomalies detected through online monitoring or following repair actions.
This approach minimizes validation overhead while preventing degraded nodes from silently re-entering
production workloads.
\begin{table*}[t]
\centering
\caption{Grey Node Classification Rates}
\label{tab:classification_error_rates}
\vskip 0.15in
\begin{small}
\begin{sc}
\begin{tabular}{lcc}
\toprule
\textbf{Metric} & \textbf{Percentage} & \textbf{Sample Size} \\
\midrule
False Positive Rate (FPR)
& 12.4\%
& 124 out of 1000 negative samples \\

False Negative Rate (FNR)
& 7.8\%
& 78 out of 1000 positive samples \\
\bottomrule
\end{tabular}
\end{sc}
\end{small}
\vskip -0.1in
\end{table*}

\begin{figure*}[t]
\centering
\begin{tikzpicture}[
  font=\small,
  node distance=10mm and 14mm,
  >=Stealth,
  block/.style={
    rectangle, rounded corners,
    draw, align=center,
    minimum height=8mm,
    minimum width=36mm
  },
  decision/.style={
    diamond, draw,
    aspect=2,
    align=center,
    inner sep=1pt
  },
  line/.style={->, thick}
]

% ===== Nodes =====
\node[block] (start) {Grey node\\ quarantined};

\node[decision, below=of start] (bad)
  {Emitting GPU / Network\\ Errors};

\node[block, right=of bad] (drivers)
  {Reboot and redeploy drivers};

\node[decision, below=of drivers] (status1)
  {Emitting GPU / Network\\ Errors};

\node[block, right=of status1] (reprov)
  {Reprovision / Re-Deploy};

\node[decision, below=of reprov] (status2)
  {Emitting GPU / Network\\ Errors};

\node[block, below=of status2] (terminate)
  {Terminate / Replace\\ node};

\node[block, left=of status2] (sweep)
  {Return for\\ node sweep};

% ===== Edges =====

% Start -> first decision
\draw[line] (start) -- (bad);

% First decision
\draw[line] (bad.east) -- node[above]{Yes} (drivers.west);
\draw[line] (bad.south) -- ++(0,-8mm) |- node[left]{No} (terminate.west);

% After driver redeploy
\draw[line] (drivers) -- (status1);

% Second decision
\draw[line] (status1.east) -- node[above]{Yes} (reprov.west);
\draw[line] (status1.south) -- node[right]{No} (sweep.north);

% After reprovision
\draw[line] (reprov) -- (status2);

% Final decision
\draw[line] (status2.south) -- node[right]{Yes} (terminate.north);
\draw[line] (status2.west) -- node[above]{No} (sweep.east);

\end{tikzpicture}

\caption{Bad Node Remediation Workflow}
\label{fig:Bad Node Remediation Workflow}
\end{figure*}
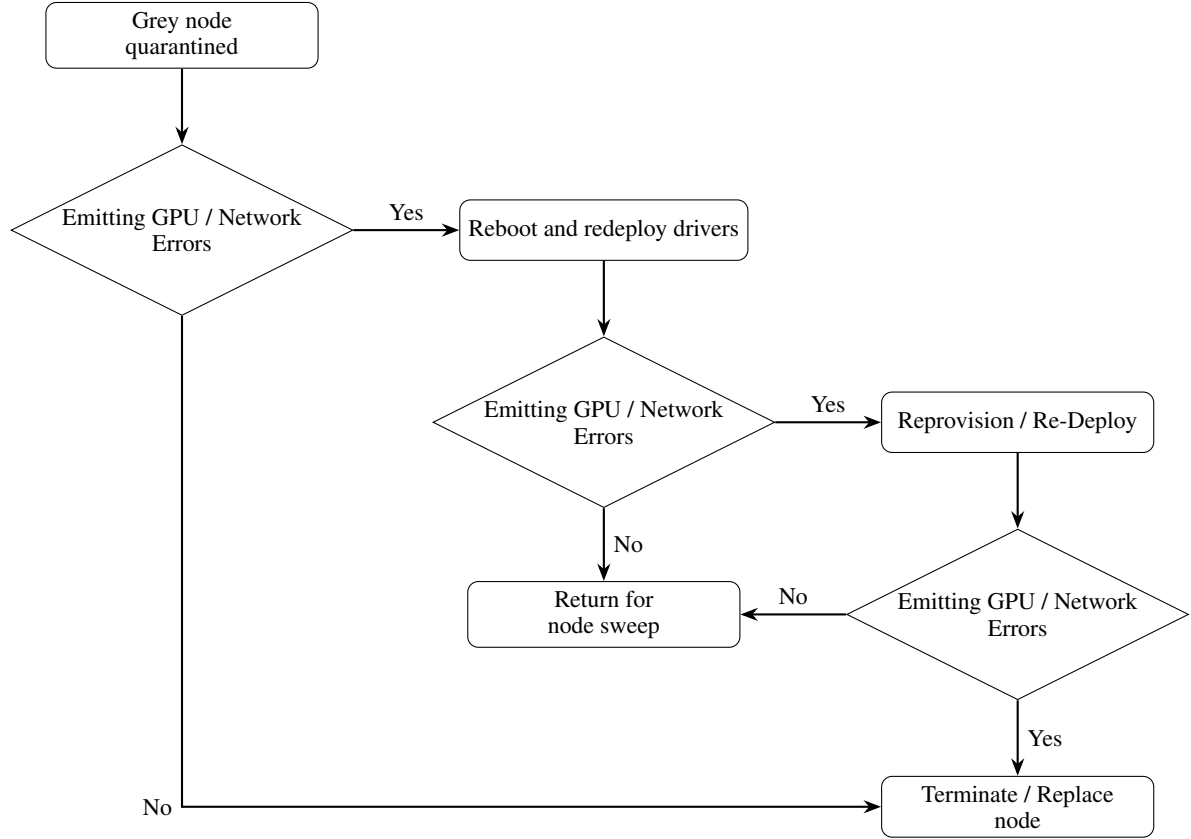

\section{Detection Parameters and Node Triage Workflow}
Given the typical fixed compute budget that clusters exhibit, node replacement is a key part to achieving a high utilization. With traditional fail-stop faults, the remediation process usually involves replacing hardware, however given that grey node failures can have multiple root causes that are not immediately visible, a more detailed triage workflow is needed. To triage grey nodes, we created a multi tiered triage workflow that can be easily automated and is focused on driving down wasted compute. The workflow exists as part of the above mentioned automated node health management workflow, specifically within the repair/mitigation section. One important note is that if a specific node was repeatedly being added to triage workflow within a certain time-period, a human manually marked it for termination without going through the entire triage process. Through the empirical observations we determined that three times within a week was sufficient evidence that a node was terminally bad and required replacement, however this may change depending on cluster configurations.

% \caption{Sample Node Triage Workflow}
\label{fig:node_triage_workflow}

\subsection{False Positive and False Negative Rates} 
The effectiveness of the grey node detection and triage workflow is highly dependent on the accuracy of the underlying classification signals used to trigger remediation actions. False positives result in healthy nodes being unnecessarily removed from service and subjected to remediation or replacement, while false negatives allow degraded nodes to continue consuming compute and negatively impacting workload performance. As such, both error modes represent a direct cost to overall cluster efficiency and must be carefully balanced.

In practice, these error rates informed the design of the tiered triage workflow shown in Table~\ref{tab:classification_error_rates}, where early remediation stages are intentionally lightweight and reversible. By combining moderately conservative detection thresholds with staged escalation and explicit health re-evaluation, the system limits the operational impact of misclassification while maintaining strong protection against sustained performance degradation.

\subsection{Sample Node Triage Workflow}
Figure~\ref{fig:Bad Node Remediation Workflow} shows a simplified node triage workflow that attempts to remediate grey nodes in a tiered manner, beginning with early termination for nodes that do not exhibit actionable error signals and progressively escalating remediation for those that do. Nodes emitting GPU or network errors are subjected to increasingly invasive recovery steps, with health checks after each stage determining whether the node can safely return to the general sweep pipeline. The workflow ultimately resolves in one of two terminal states, either returning the node for further validation once errors clear or terminating and replacing the node when repeated remediation fails, thereby limiting wasted compute and operational churn.

\section{Results}

\subsection{Cluster Setup}

Our experiments were conducted on a high-performance GPU cluster equipped with thousands of GPUs deployed in a traditional 8 GPU node setup. Intra-node communication is handled by NVLink with inter-node communication handled by external network cards. Hardware metrics for the online system are collected through a combination of NVIDIA DCGM and custom sidecar processes with custom polling frequencies between 30 seconds and 1 minute. The cluster supports both long-term foundation model pretraining and short-term post-training and inference workloads. The interconnect fabric is based on a high-bandwidth network card, which enables high-throughput and low-latency communication across nodes. 

The software environment is built on a distributed training framework with optimized communication primitives, supporting mixed-precision training with BF16 and model parallelism strategies such as data, tensor, pipeline and expert parallelism. 

\subsection{Cluster-Level Results}

\begin{table*}[t]
\centering
\caption{Ablation study of system components}
\label{tab:ablation_study}
\vspace{4pt}
\small
\begin{tabular}{
>{\centering\arraybackslash}m{5.0cm}
>{\centering\arraybackslash}m{2.5cm}
>{\centering\arraybackslash}m{3.2cm}
>{\centering\arraybackslash}m{2.0cm}
>{\centering\arraybackslash}m{3.2cm}
}
\toprule
\textbf{Method} &
\textbf{Avg. MTTF (h)} &
\textbf{Avg. Human Interval (h)} &
\textbf{Avg. MFU} &
\textbf{Detects HW Degradation} \\
\midrule
NCCL / Burnin Tests Only
& 6.6 & 5.6 & 5\% & No \\

NCCL / Burnin + Node Sweep
& 8.1 & 2.0 & 10\% & No \\

NCCL / Burnin + Online Monitoring + Node Sweep
& 9.2 & 1.2 & 14\% & Yes \\

NCCL / Burnin + Online Monitoring + Enhanced Node Sweep
& \textbf{16.7} & \textbf{0.5} & \textbf{17\%} & Yes \\
\bottomrule
\end{tabular}
\end{table*}

% \begin{table*}[t]
% \centering
% \caption{Cluster-level reliability metrics before and after optimization.}
% \label{tab:cluster_reliability}
% \vskip 0.15in
%   \centering
%   \begin{small}
%   \begin{sc}
% \begin{tabular}{lcc}
% \toprule
% \textbf{Metric} & \textbf{Before Optimization} & \textbf{After Optimization} \\
% \midrule
% Mean Time to Failure (MTTF)  & 4.73 hours & 16.7 hours \\
% Human Intervention Interval & 2.0 hours & 0.5 hours \\
% Mean FLOPS Utilization (MFU) & 10\% & 17\% \\
% \bottomrule
% \end{tabular}
% \end{sc}
%   \end{small}
%   \vskip -0.1in
% \end{table*}

\begin{figure}[t]
    \centering

    % 第一张图
    \includegraphics[width=0.9\linewidth]{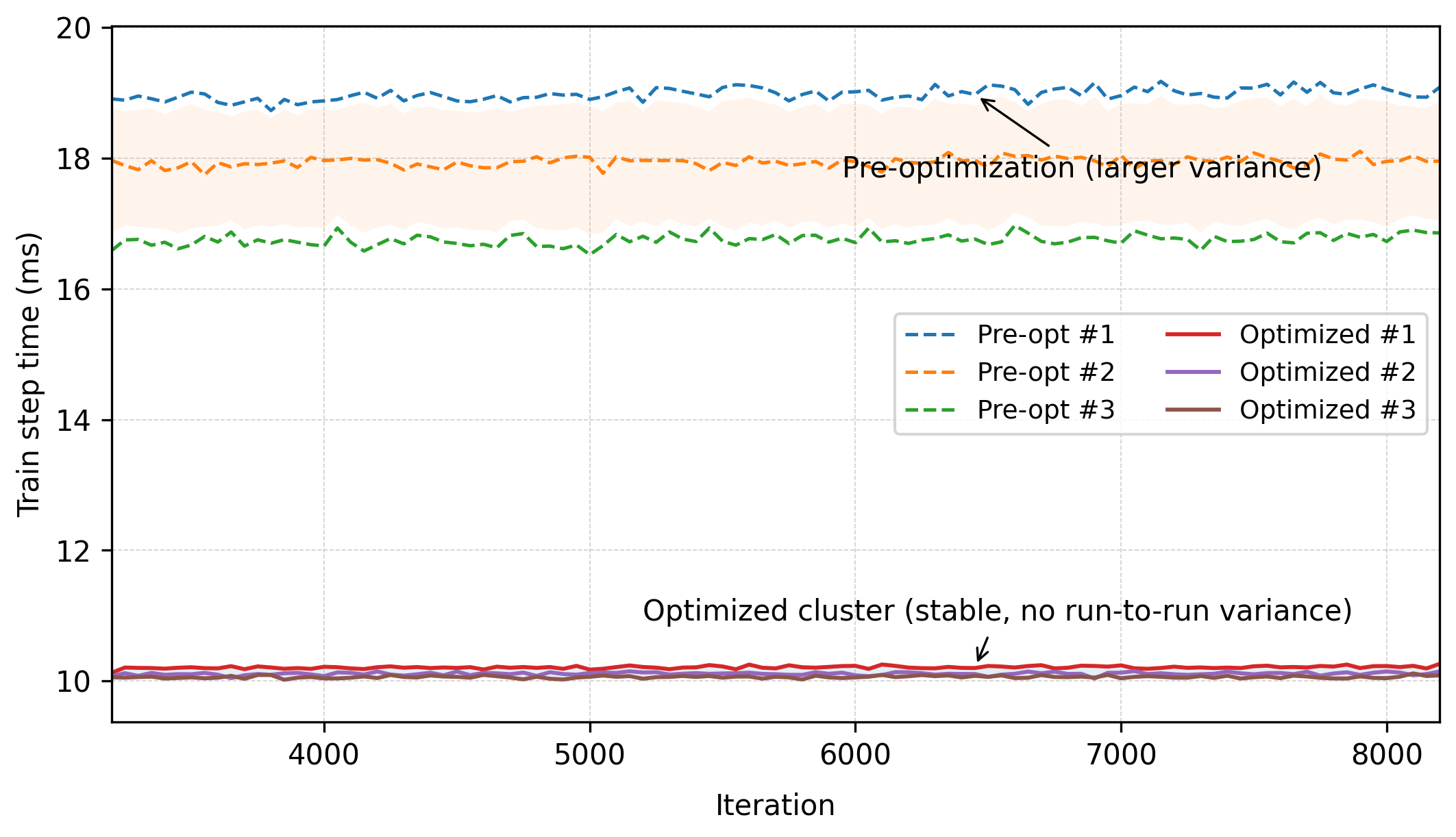}
    \caption{Comparison of run-to-run variance in training step time before and after applying the proposed node health monitoring system.}
    \label{fig:train_step_variance}

    \vspace{2pt}

    % 第二张图
    \includegraphics[width=0.9\linewidth]{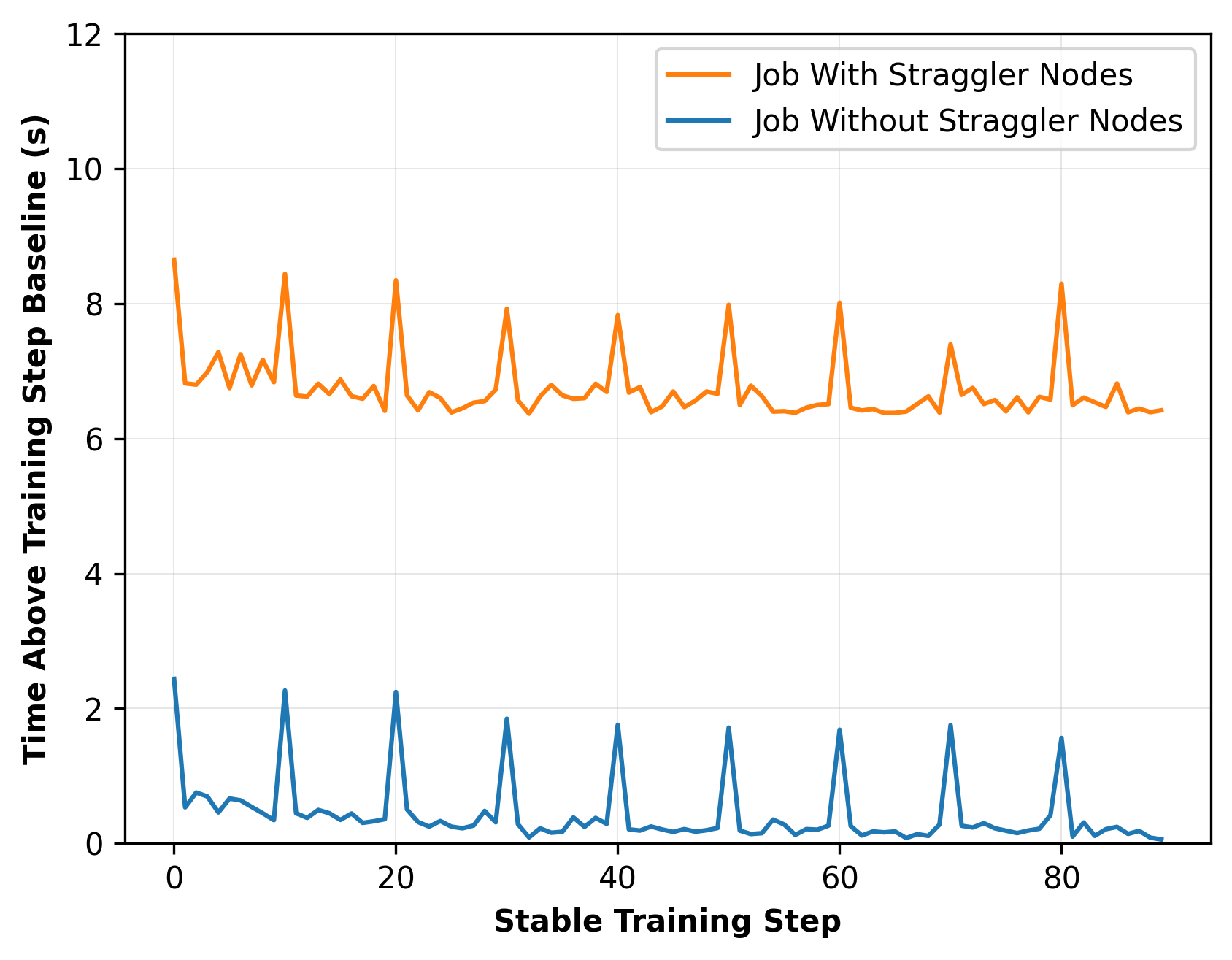}
    \caption{Training step time reduction of 7 seconds after applying node health monitoring and selection strategies.}
    \label{fig:train_step_reduction}
\end{figure}

The main experimental results are derived from large-scale foundation model pretraining conducted over several months across thousands of GPUs. The model was trained on a diverse corpus containing trillions of tokens spanning multiple domains. We evaluate the proposed system through both quantitative metrics and qualitative observations to assess its impact on large-scale training efficiency and reliability. Qualitatively, the system maintains high reliability during prolonged pretraining runs. Even under large-scale workloads, faulty nodes are effectively detected, isolated, and recovered through the combined online monitoring and offline node sweep workflow, minimizing manual maintenance overhead.Furthermore, as illustrated in Figure~\ref{fig:train_step_variance}, the run-to-run variance in training step time is reduced from 20\% to only 1\%, indicating significantly more stable cluster behavior.

In addition to improved stability, the proposed health monitoring and node selection strategies significantly reduce training latency. As shown in Figure~\ref{fig:train_step_reduction}, the average training step time decreases from 17 seconds to 10 seconds, saving approximately 7 seconds per step and resulting in an overall \textbf{70\% improvement in training efficiency}.

To better understand the contribution of each component, we further conduct an ablation study that incrementally enables online monitoring and offline node sweep As summarized in Table~\ref{tab:ablation_study}. Enabling online monitoring alone improves the average MTTF by 14\% (from 8.1h to 9.2h), reduces the average human intervention interval by 40\% (from 2.0h to 1.2h), and increases MFU from 10\% to 14\%. Introducing the enhanced offline node sweep yields an additional 82\% improvement in average MTTF (from 9.2h to 16.7h), further reduces human intervention time by 58\% (from 1.2h to 0.5h), and increases MFU from 14\% to 17\%.

These results demonstrate that the combination of online monitoring and offline node sweep is essential for identifying both acute failures and long-running fail-slow nodes whose degradations may not manifest as explicit hardware errors but can significantly impact training step time. Excluding such nodes accounts for a substantial portion of the MFU and MTTF improvements observed at scale.

\subsection{Implications at Larger Scale}

As training scale increases, the impact of grey nodes on overall efficiency becomes increasingly
amplified.
In large-scale deployments with thousands of GPUs and frequent synchronization points, even minor
node-level performance degradation can propagate through collective operations and lead to
disproportionate slowdowns at the job level.
This amplification effect makes timely detection and mitigation of grey nodes substantially more
critical at larger scales.

Importantly, the proposed online monitoring and offline node sweep mechanisms are designed to remain
effective as system scale grows.
The online component continuously operates with low overhead across the cluster, enabling scalable
detection of performance anomalies, while the offline node sweep validates isolated nodes
independently of overall cluster size.
Together, this design allows the framework to scale to larger training systems while maintaining
stable performance and manageable operational cost.

\section{Conclusion}
This paper presents a practical, production-grade system for node health monitoring and straggler detection in large-scale foundation model training. The proposed system continuously monitors critical hardware and network signals during pretraining across thousands of GPUs, and complements this online monitoring with an efficient offline node-sweep component to systematically identify performance-degrading nodes, including edge cases that traditional NCCL tests and GPU burn-in procedures often fail to expose.

Through extensive evaluations on real-world, production-scale training workloads, the system achieves up to 70\% improvement in training step efficiency, reduces run-to-run performance variance from 20\% to 1\%, and substantially improves overall cluster reliability and utilization. These results demonstrate that a system-level, closed-loop approach—combining continuous monitoring with targeted validation—can effectively mitigate fail-slow behaviors and significantly enhance both the performance and operational stability of large-scale GPU clusters. The proposed system is scalable, lightweight, and readily deployable in modern foundation model training environments.

\bibliographystyle{mlsys2025}
\bibliography{example_paper}

\end{document}